\documentclass[11pt]{article}
\usepackage{xspace}
\usepackage{graphicx}
\usepackage{amsmath}
\usepackage{amssymb}
\usepackage{color}
\usepackage{textcomp}
\usepackage{caption2}

\definecolor{Red}{rgb}{1,0,0}
\definecolor{Green}{rgb}{0,1,0}
\definecolor{Blue}{rgb}{0,0,1}
\definecolor{Black}{rgb}{0,0,0}

\def\beq{\begin{equation}}
\def\eeq#1{\label{#1}\end{equation}}
\def\eeqn{\end{equation}}

\def\beqa{\begin{eqnarray}}
\def\eeqa#1{\label{#1}\end{eqnarray}}
\def\eeqan{\end{eqnarray}}

\let\bar=\overbar

\def\Dslash{\not{\hbox{\kern-4pt $D$}}}
\def\dslash{\not{\hbox{\kern-2pt $\del$}}}

\def\msb{{\bar{\ssstyle M \kern -1pt S}}}

\def\Title#1{\begin{center} {\Large {\bf #1} } \end{center}}

\begin{document}

\Title{The ANTARES Neutrino Telescope}

\bigskip
\bigskip


\begin{raggedright}  

{\it Chiara Perrina\index{Perrina, C.} on behalf of the ANTARES Collaboration\\
``La Sapienza'' University of Roma and INFN\\
Piazzale Aldo Moro, 5\\
00185 Roma, Italy}\\
\end{raggedright}
\vspace{1.cm}

{\small
\begin{flushleft}
\emph{To appear in the proceedings of the Prospects in Neutrino Physics Conference, 15 -- 17 December, 2014, held at Queen Mary University of London, UK.}
\end{flushleft}
}
\section{Introduction}

Great progress has been made in the last decades for what concerns the energy spectrum and the composition of cosmic rays but their origin remains uncertain. Neutrinos are a valid probe for the exploration of the high-energy sky, since they provide complementarity with photons and cosmic rays observations, and even enlarge the amount of collected information, as they can emerge from dense media and travel across cosmological distances without being deflected by magnetic fields nor absorbed by ambient matter and radiation. Hadronic interaction models predict high-energy ($>$ TeV) neutrinos from a wide range of astrophysical objects, from galactic sources such as Supernovae Remnants or Microquasars to the most powerful extragalactic emitters such as Active Galactic Nuclei and Gamma-Ray Bursts~\cite{Becker:2007sv}. The weakness of neutrino interactions and the faintness of the expected cosmic neutrino fluxes require large volumes of transparent medium (water or ice) instrumented with arrays of photosensors, in order to detect the Cherenkov light induced by the charged leptons produced in the neutrino interactions with matter in or around the detector. Although the muon identification represents the most straightforward detection channel, showers induced by electron and tau neutrinos can also be detected. Neutrino Telescopes are installed at great depths and optimized to detect up-going muons produced by neutrinos that have traversed the Earth, in order to limit the background from down-going atmospheric muons. Atmospheric neutrinos, with an energy spectrum $\propto E_{\nu}^{-3}$, traverse the Earth and interact close to the detector representing an irreducible background. Neutrinos of cosmic origin which are expected to have an energy spectrum $\propto E_{\nu}^{-\alpha}$ with $\alpha = 1$ or $2$ can be identified only on a statistical basis.
\section{The ANTARES detector}

ANTARES (Astronomy with a Neutrino Telescope and Abyss environmental RESearch) is the first undersea Neutrino Telescope and the only one currently operating~\cite{Collaboration:2011nsa}. It is located at a depth of 2475 m in the Mediterranean Sea, $\sim 40 $ km off the French coast south-east Toulon. It consists of a matrix of 885 photomultiplier tubes (PMTs) arranged into 12 strings anchored to the sea bed and maintained vertical by buoys, connected to a junction box which distributes the electrical power and transmits the data to shore through an electro-cable. The PMTs are orientated at $45^\circ$ downwards in order to maximize the sensitivity to Cherenkov light from up-going muons. See~\cite{AdrianMartinez:2012ky,Aguilar:2010aa} for details on the detector. The median angular resolution achieved for muon tracks, assuming an energy spectrum $\propto E_{\nu}^{-2}$, is $< 0.4^\circ$ allowing good performance in the searches for neutrino point sources. Its location in the Northern Hemisphere allows for surveying a large part of the Galactic Plane, including the Galactic Centre, thus complementing the sky coverage of the IceCube detector installed at the South Pole.
\section{Searches for diffuse neutrino fluxes}
\begin{figure}[]
\begin{center}
\includegraphics[width=0.4\columnwidth]{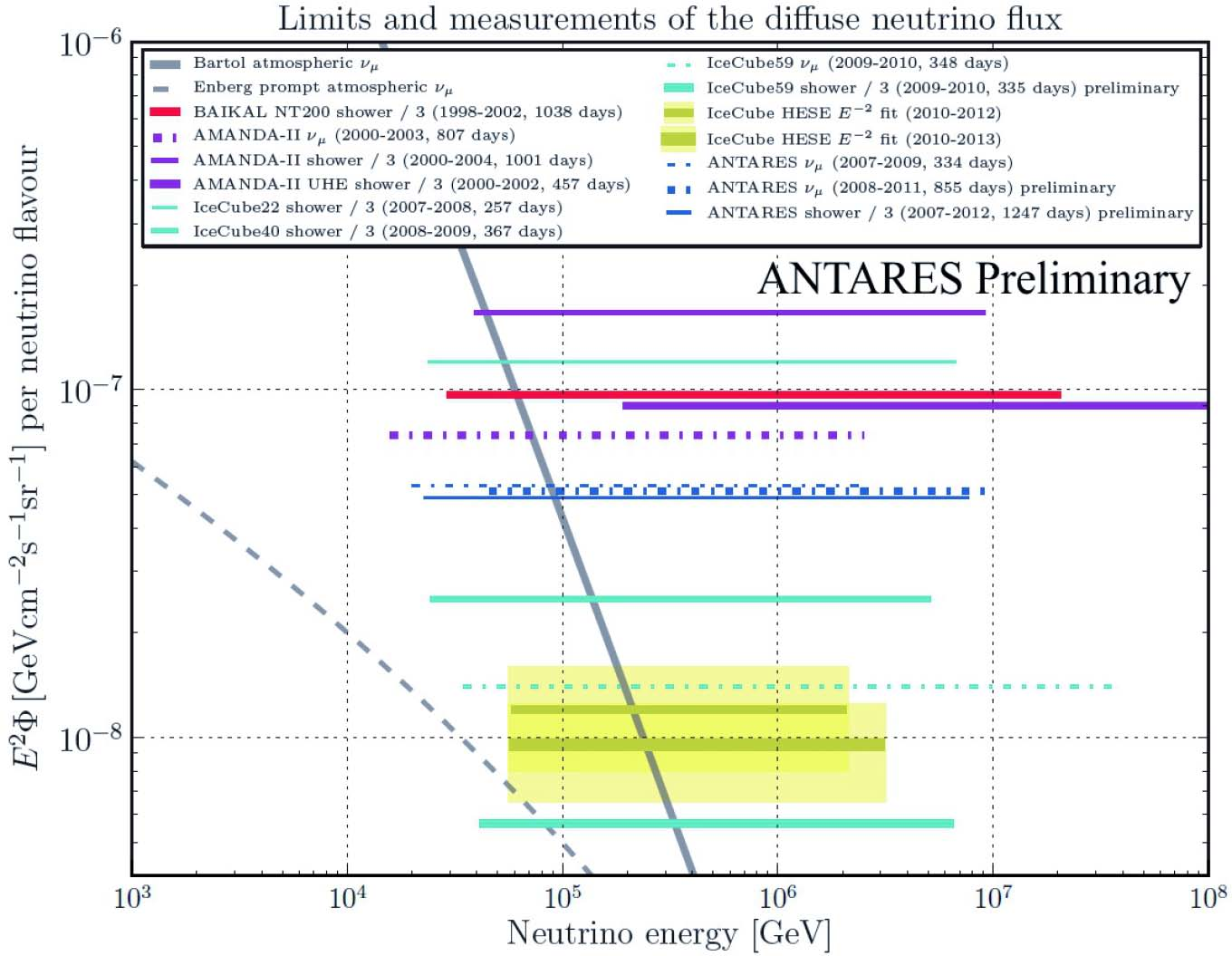}
\caption{$90\%$ C.L. upper limits on an $E_{\nu}^{-2}$ diffuse neutrino flux for different experiments and analyses. The ANTARES upper limits set by the muon neutrino tracks and shower events analyses are represented by blue lines.}
\label{fig:diffuse_limits_and_measurements}
\end{center}
\vspace{-0.7cm}
\end{figure}

A search for a cosmic neutrino diffuse flux using up-going muon neutrino events has been performed using the data collected from December 2007 to December 2011 (live-time = 855 days). The first step, in order to reduce the background, consists in applying loose cuts on the muon track reconstruction quality parameter, called $\Lambda$, and on the angular error estimate obtained by the reconstruction fit. Then an optimisation procedure based on the Model Rejection Factor (MRF) method has been applied with the aim to determine the energy cut yielding the best sensitivity. The number of events passing the entire selection is 8 while 8.4 events are expected as background. This result yields a $90\%$ confidence level (C.L.) upper limit on a diffuse cosmic signal of
\[E_{\nu}^{2}\phi_{90\%} = 5.1 \times 10^{-8} \text{ GeV} \text{ cm}^{-2}\text{ s}^{-1}\text{ sr}^{-1}\]
for $E_{\nu}$ in the range 45 TeV - 10 PeV (Fig. \ref{fig:diffuse_limits_and_measurements}). 

The search for a diffuse cosmic neutrino flux has also been performed using shower events, for the February 2007 - December 2012 data sample (live-time = 1247 days). Once the shower is reconstructed a series of selection criteria have been imposed. First, a muon filter is used to reject events compatible with a muon track, then the event is required to be up-going and with energy higher than 10 TeV. After these selection criteria, the number of expected background events is 4.9 and the number of the observed ones is 8. This excess is interpreted as a background fluctuation and the following $90\%$ C.L. upper limit is derived:
\[E_{\nu}^{2}\phi_{90\%} = 4.9 \times 10^{-8} \text{ GeV} \text{ cm}^{-2}\text{ s}^{-1}\text{ sr}^{-1}\]
for $E_{\nu}$ in the range 23 TeV - 7.8 PeV (Fig. \ref{fig:diffuse_limits_and_measurements}).

\section{Searches for point-like neutrino sources}
\begin{figure}[!h]
\begin{center}
\includegraphics[width=0.3\columnwidth]{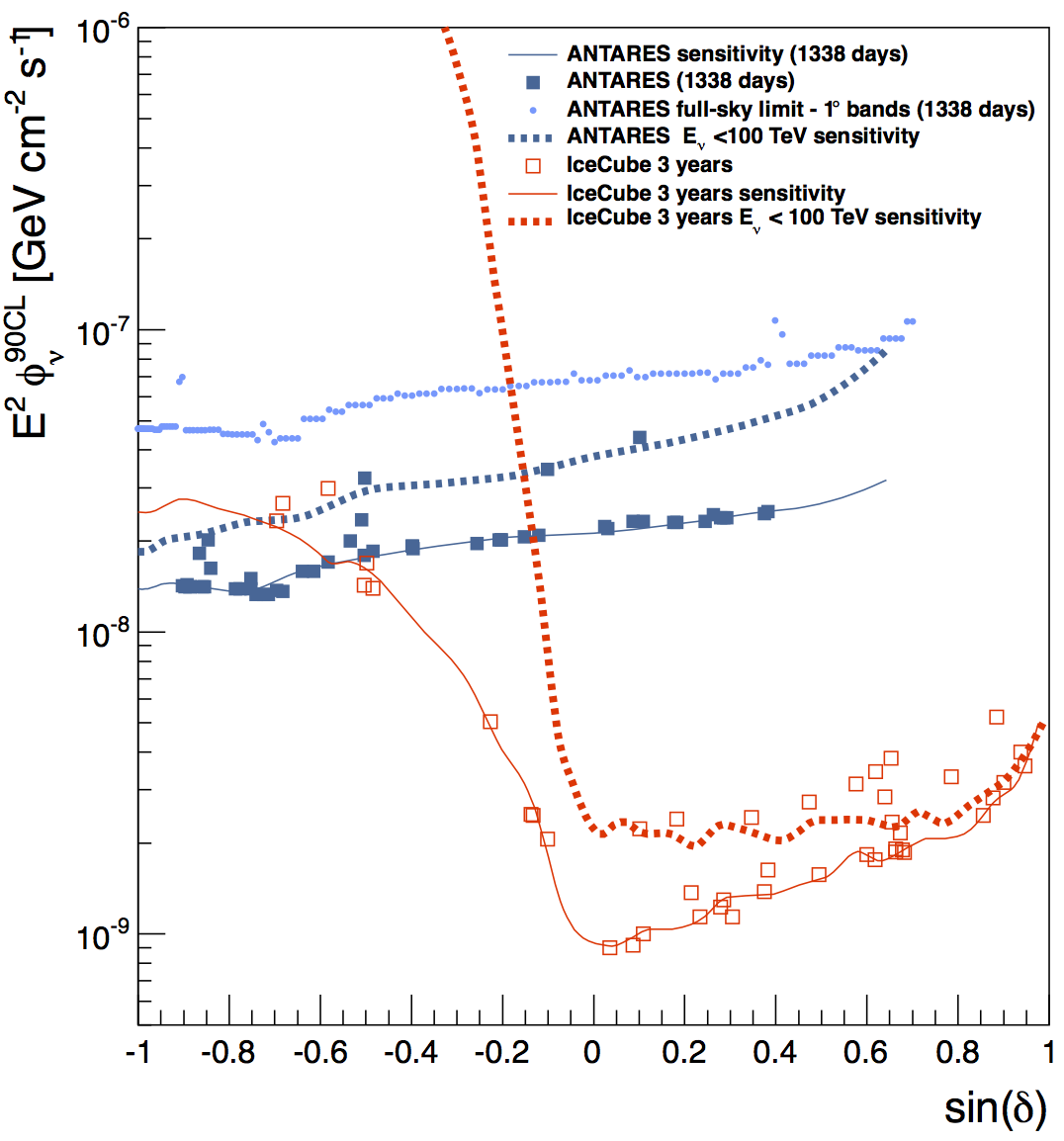}
\caption{$90\%$ C.L. flux upper limits and sensitivities for a point-source with an $E_{\nu}^{-2}$ spectrum as a function of the declination, for six years of ANTARES data. IceCube results are also shown for comparison~\cite{Aartsen:2013uuv}. The dashed dark blue (red) line indicates the ANTARES (IceCube) sensitivity for neutrino energies lower than 100 TeV, showing that the IceCube sensitivity for sources in the Southern Hemisphere is mostly due to events of higher energy (taken from~\cite{Adrian-Martinez:2014wzf}).}
\label{fig:point_upper_limits_and_sensitivities}
\end{center}
\vspace{-0.5cm}
\end{figure}

The detection of point-like sources is possible by identifying a significant excess of events from particular spots (or small regions) of the sky. ANTARES has performed these searches~\cite{Adrian-Martinez:2014wzf} on the 2007-2012 data sample (live-time = 1338 days). The final neutrino sample has been obtained after tight cuts on the reconstruction quality parameter $\Lambda$, on the estimated angular resolution ($< 1^\circ$) and on the zenith angle ($⁡\cos \theta < 0.1$). It contains 5516 neutrino candidates, with a predicted atmospheric neutrino purity of $\sim 90\%$. Two different searches have been performed: 
\begin{enumerate}
\item a time-integrated full-sky search looking for an excess of events over the atmospheric neutrino background in the declination range [-90$^\circ$, +48$^\circ$];
\item a candidate-list search looking for events in the directions of a predefined list of 50 candidate sources of interest which are known gamma-ray emitters and potential sites for hadronic acceleration.
\end{enumerate}
In both searches no significant excess over the background has been found and upper limits have been derived (Fig. \ref{fig:point_upper_limits_and_sensitivities}). 

\begin{figure}[]
\begin{center}
\includegraphics[width=0.6\columnwidth]{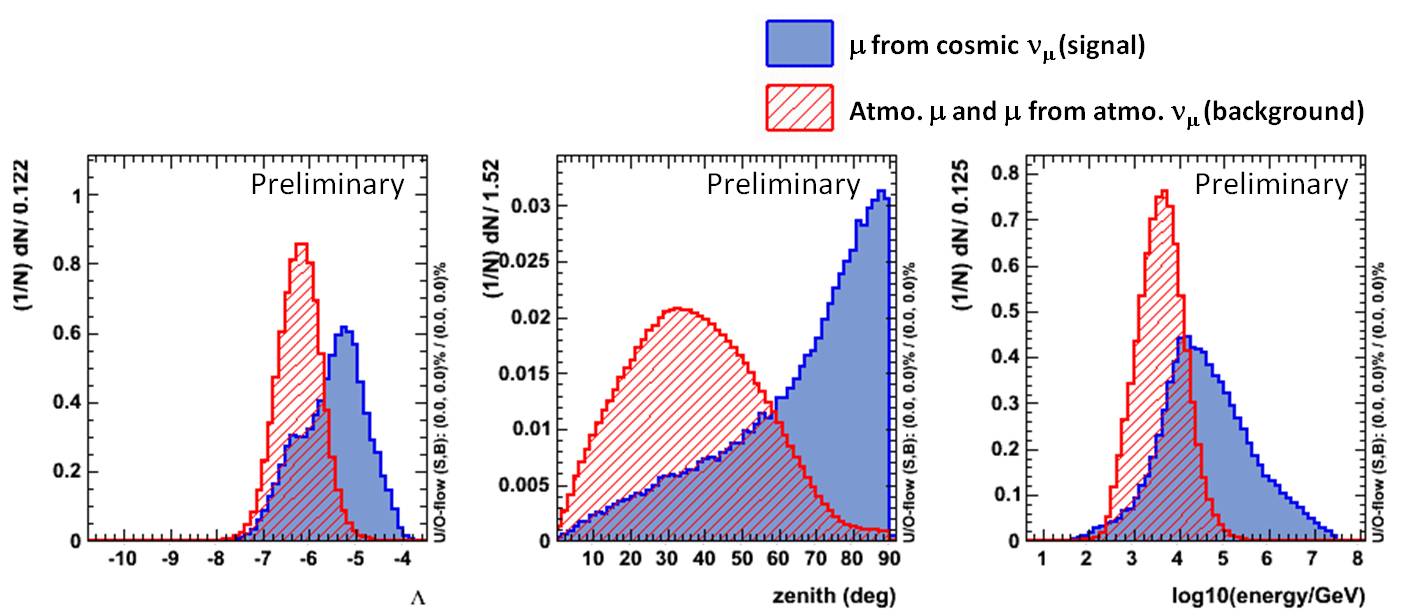}
\caption{Distribution of $\Lambda$, zenith angle and energy for Montecarlo-generated down-going events in ANTARES, comparing expected signal and atmospheric background.}
\label{fig:down-going_variables}
\end{center}
\vspace{-1.0cm}
\end{figure}

Until now the point-like sources search has been restricted to the Southern Hemisphere: only up-going events have been studied in order to reject the atmospheric muons background from cosmic-ray-induced showers. The development of a strategy for the identification of down-going neutrino events is going on. The goal is to extend the field of view of ANTARES and increase the energy threshold of the search. In such analysis three important variables have to be considered: the energy of the event, its coming direction and the track reconstruction quality ($\Lambda$). The best signal/background separation can be obtained by looking for high energy, mostly horizontal (zenith $\sim 90^\circ$) and well reconstructed events (Fig. \ref{fig:down-going_variables}). 

\section{Summary}
The ANTARES neutrino telescope is in its $8^{\text{th}}$ year of operation. Thanks to its location and to the excellent angular resolution it is yielding diffuse neutrino flux sensitivity in the relevant energy range and the best limits in the world for many (galactic) sources in the Southern Hemisphere, especially for $E_{\nu} < 100$ TeV. Beyond Astrophysics, searches for dark matter and exotic particles are also performed. More competitive results are expected in the future as ANTARES will continue taking data at least until the end of 2016, when it will give way to the next-generation KM3NeT detector.

%

\end{document}